\documentstyle[12pt,psfig,epsf]{article}

\def\bc{\begin{center}}
\def\ec{\end{center}}
\def\beq{\begin{equation}}
\def\eeq{\end{equation}}

\def\ksi{\xi}

\def\F#1{Fig.~\ref{f#1}}
\def\lx1{x_1=\xi\rightarrow nx_0}
\def\x#1{x_#1\rightarrow 0}
\def\k{{\bf k}}
\def\n{\overline{n}}

\def\v#1#2#3#4{
\put(#1,#2){\circle*{2.00}}
\put(#1,#3){\circle*{2.00}}
\put(#1,#2){\dashbox{1.5}(0,#4)[t]{}}
}

\begin{document}
\thispagestyle{empty}

\bc
{\large CUMULATIVE PHENOMENA THROUGH\\
THE QUARK-PARTON DIAGRAM SUMMATION\\
\vskip 1 mm
AT THRESHOLDS}
\ec

\bc
{\large M. A. Braun and V. V. Vechernin
\footnote{E-mail: vecherni@snoopy.niif.spb.su}}\\
{\it Department of Theoretical Physics,
St.Petersburg State University}\\
{\it 198904 St.Petersburg, Russia}
\ec

\medskip

\begin{abstract}
A microscopic treatment of cumulative phenomena based on
perturbative QCD calculations of the corresponding quark diagrams
near the threshold is presented.
To sum all diagrams like these
the special technique based on the
recurrence relations was developed.
The $x$-behaviour of the nuclear structure function $F_{2}(x)$
in the cumulative region $x>1$
was found to be roughly exponential, governed by an effective
coupling constant, which depends on the QCD coupling constant
and quark mass.
Two mechanisms for cumulative
particle production, direct and spectator ones, were analysed. It was
shown that due to final state interactions the leading terms of the direct
mechanism contribution are cancelled and the spectator mechanism is the
dominant one.
It leads to a smaller slope of the particle
production rates compared to the slope of the nuclear structure function
in the cumulative region,
in agreement with the recent experimental data.
The slope difference is due to additional multiple interactions
between nuclear and projectile partons which
enter the spectator mechanism for the cumulative production.
The different versions of hadronization mechanisms of the produced
cumulative quarks into hadrons are also
discussed.

\end{abstract}

\medskip

\section{Introduction}

One of the most interesting fragmentation processes is
the production of fast hadrons in $hA$ interactions
in the nuclear fragmentation region with the value of the
Feynman scaling variable: $x>1$,
i.e. in the region in which the production of this fast hadron
is kinematically prohibited for the $hN$ interaction.
Here $x$ is the longitudinal momentum per
nucleon of the fast moving nucleus carried
away by the produced particle.
It means that a produced particle
carries a momentum much greater than the average one of a constituent
nucleon.
The nuclear fragmentation processes like these usually
are referred as the cumulative ones.
For the deep inelastic $lA$ scattering
the cumulative region is $x>1$ in terms of
Bjorken scaling variable defined for individual $lN$ interactions.

The theoretical investigations of the nuclear fragmentation processes
in this cumulative region are of great physical importance
as they enable
to study multi-nucleon short-range correlations
in nuclei (so called fluctons)
and
to get the information on
the high-density nuclear matter clusters
which always are being in nucleus.

From the modern point of view
these compact multi-quark dense hadronic matter clusters in nuclei
can be also considered as
the quark-gluon plasma clusters at zero temperature
which investigation is of interest also in the light
of recent efforts to detect the indications
of high temperature quark-gluon plasma formation
in heavy-ion collisions on colliders at super high energies.

An adequate treatment of the cumulative effect requires the quark
language \cite{7,8}.
A phenomenological treatment of cumulative phenomena based on nuclear
quark distributions and QCD evolution equations was proposed in \cite{8}.
It uses an {\it ad hoc} relation between the structure functions
and particle production rate, based on the nucleonic (or fluctonic) picture
of the nucleus rather than the quark one. Starting from the quark picture of
hadrons the famous quark counting rules have been proposed \cite{11,12},
which,
in particular, describe the threshold behaviour of the structure functions.
In \cite{12} it was noted that these rules also apply to nuclei at the deep
cumulative threshold $x\rightarrow A$. In \cite{13,14} it was stressed
  that the cumulative
phenomena involve a particular type of contributions,
 the so-called intrinsic
diagrams, in which several partons of the nucleus cooperate to  determine the
intrinsic hardness of the nuclear wave function.

However the quark counting rules alone do not allow to find the cumulative
probabilities. To calculate these, one needs besides to know the coefficients
of the quark counting rules behaviour, which determine the relative weights
with which different contributions combine to give the total probability, as
well as the overall  normalization.
Recently we have proposed \cite{NPB,dub95} a microscopic model for the
theoretical description of the cumulative
phenomena based on perturbative QCD calculations of the corresponding
quark-parton Feynman diagrams near the thresholds
and calculated these coefficients
in the framework of the proposed model.
By means of the specially developed
technique based on the recurrence relations
all diagrams of that kind were summed.
As a result, we obtain formulas for the nuclear
structure function and particle production rate at $x>1$, which allow direct
comparison with experiment.
The proposed method of the summation of intrinsic diagrams at
cumulative thresholds
based on the recurrence relation technique
are also discussed in the present paper.

\section{General formulae}
\subsection{The nuclear structure function at $x>1$}

Consider the virtual $\gamma$-nucleus scattering cross-section in the
system, where the nucleus is moving fast along the $z$-axis
with the longitudinal momentum per nucleon $p_{z}$.
As a reasonable first approximation, we may treat the nucleus as a
colllection of $N=3A$ valence quarks, which, on the average, carry each
momentum $x_{0}p_{z}$ with $x_{0}=1/3$.
In the following we assume that the quark longitudinal momentum
distribution is sharp enough, so that in all places, except in the quark
wave function itself, we can safely put the initial quark longitudinal
momentum equal to $x_{0}p_{z}$. The probing $\gamma$ quantum will see a
"active" quark with its momentum much greater
than $x_{0}p_{z}$ only if this quark has
 interacted with several other quarks ("donors") and has taken some of
 their longitudinal momenta. The maximum possible value of the
 longitudinal momentum thus
accumulated  is $Nx_{0}p_{z}=Ap_{z}$. It corresponds to $N-1$
interactions with all
other quarks, whose longitudinal momentum has become equal to zero each.
It is well-known that interactions which make the longitudinal momentum
of one of the quark equal to zero may be treated by perturbation theory
\cite{14}.
These considerations were used in numerous
calculations of hadronic structure functions near $x=1$, which revealed
that the results depend heavily on the quark distribution in the initial
hadron (see e.g \cite{15}). In the case of a nucleus the calculations in
principle remain the same. With a number of quarks considerably larger,
their distribution in $x$ evidently becomes much narrower, so that we
may hope that the results would be less sensitive to its form nor to its
spin-colour structure. On the other hand the number of relevant diagrams
rises tremendously. Their summation presents the main techical difficulty
in the nuclear case.

%Fig.1
\begin{figure}

\unitlength=0.5mm
\linethickness{0.4pt}
\begin{picture}(90.00,100.00)(-80,20)
\put(20.00,41.00){\line(1,0){15.00}}
\put(20.00,39.00){\line(1,0){15.00}}
\put(40.00,40.50){\oval(10.00,41.00)[]}
\put(45.00,24.00){\line(1,0){30.00}}
\put(45.00,29.00){\line(1,0){30.00}}
\put(45.00,34.00){\line(1,0){30.00}}
\put(45.00,39.00){\line(1,0){30.00}}
\put(45.00,44.00){\line(1,0){30.00}}
\put(45.00,49.00){\line(1,0){30.00}}
\put(45.00,54.00){\line(1,0){30.00}}
\put(71.00,74.00){\circle{10.00}}
\put(44.00,59.00){\line(2,1){23.00}}
\put(76.00,73.00){\line(6,-1){11.00}}
\put(87.00,71.33){\line(0,-1){0.33}}
\put(75.00,77.00){\line(6,1){12.00}}
\put(18.00,40.00){\makebox(0,0)[rc]{$Ap$}}
\put(85.50,47.50){\oval(5.00,15.00)[lb]}
\put(85.50,32.50){\oval(5.00,15.00)[lt]}
\put(80.00,33.00){\oval(6.00,28.00)[rb]}
\put(90.00,41.00){\makebox(0,0)[lc]{quarks}}
\put(80.00,48.00){\oval(6.00,16.00)[rt]}
\put(56.00,68.00){\makebox(0,0)[rb]{$k$}}
\put(76.00,75.00){\line(1,0){12.00}}
\put(45.00,100.00){\makebox(0,0)[rc]{$\gamma$}}
\put(59.00,90.00){\makebox(0,0)[lb]{$q$}}

\multiput(67,78)(-2,2){10}{\makebox(0,0)[cc]{$\circ$}}

\end{picture}
\caption{The deep inelastic $\gamma A$ scattering}
\label{f1}
\end{figure}
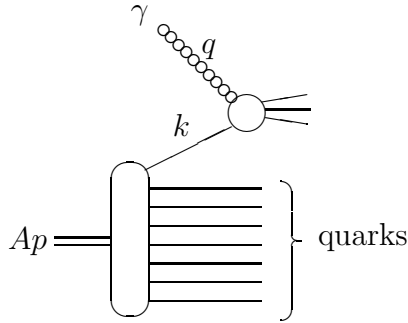

In general form the diagram
for the deep inelastic $\gamma A$ scattering
is shown in Fig.~\ref{f1}.
In the
following we use the light-cone variables $p_{\pm}=(p_{0}\pm p_{z})/
\sqrt{2}$.
The momentum
of the active quark will be denoted as $k$ and its scaling variable
 is $\ksi =k_{+}/p_{+}\simeq k_z/p_z$.
We shall study a general case when $n-1$ quarks aquire
their momenta equal to zero as a result of interquark interactions,
giving their momentum to the active quark.
 Evidently with $n-1$ momentum exchanges the $\ksi$
maximal value is $nx_{0}$. The upper blob in Fig.~\ref{f1} represents the
active
 quark's structure function. For simplicity we shall treat quarks as
 scalar particles.
 Then the contribution of the diagram of Fig.~\ref{f1} to
the nuclear structure function $F_{2}^{(A)}$ is given by
\beq
F_{2}^{A}(x,q^{2})=\int_{x} (d\ksi /\ksi)D(\ksi )f_{2}(x/\ksi,q^{2})
\label{12}
\eeq
where $D(\ksi )/\ksi$ has the meaning of the probability to find in the
nucleus a quark with the longitudinal momentum $k_z=\ksi p_{z}$,
$f_{2}$ is the quark structure function and
$x =q^{2}/2qp$.
In the cumulative region $1<x<\ksi<A$.

\subsection{Cumulative particle production}

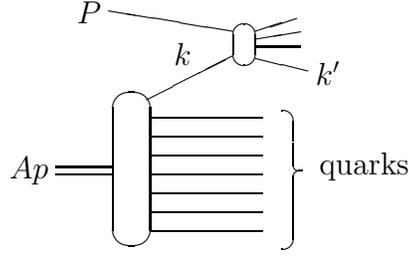
\begin{figure}

\unitlength=0.5mm
\linethickness{0.4pt}
\begin{picture}(90.00,82.33)(-70,20)
\put(20.00,41.00){\line(1,0){15.00}}
\put(20.00,39.00){\line(1,0){15.00}}
\put(40.00,40.50){\oval(10.00,41.00)[]}
\put(45.00,24.00){\line(1,0){30.00}}
\put(45.00,29.00){\line(1,0){30.00}}
\put(45.00,34.00){\line(1,0){30.00}}
\put(45.00,39.00){\line(1,0){30.00}}
\put(45.00,44.00){\line(1,0){30.00}}
\put(45.00,49.00){\line(1,0){30.00}}
\put(45.00,54.00){\line(1,0){30.00}}
\put(44.00,59.00){\line(2,1){23.00}}
\put(18.00,40.00){\makebox(0,0)[rc]{$Ap$}}
\put(85.50,47.50){\oval(5.00,15.00)[lb]}
\put(85.50,32.50){\oval(5.00,15.00)[lt]}
\put(80.00,33.00){\oval(6.00,28.00)[rb]}
\put(90.00,41.00){\makebox(0,0)[lc]{quarks}}
\put(80.00,48.00){\oval(6.00,16.00)[rt]}
\put(56.00,68.00){\makebox(0,0)[rb]{$k$}}
\put(70.00,73.50){\oval(6.00,11.00)[]}
\put(67.00,77.00){\line(-6,1){33.00}}
\put(32.00,82.00){\makebox(0,0)[rc]{$P$}}
\put(73.00,70.00){\line(4,-1){14.00}}
\put(73.00,73.00){\line(1,0){12.00}}
\put(73.00,75.00){\line(6,1){12.00}}
\put(73.00,77.00){\line(3,1){11.00}}
\put(89.00,66.00){\makebox(0,0)[lc]{$k'$}}
\end{picture}

\caption{Cumulative particle production, the direct mechanism.}
\label{f2}

\end{figure}

We now turn to the production of cumulative particles which have $x>1$
in the system where the nucleus is moving fast along the $z$-
axis. All the contributions can be divided into a direct part, in which
the projectile interacts with the created cumulative quark (Fig.~\ref{f2})
and which is a straightforward generalization of Fig.~\ref{f1}, and a
spectator part in which the projectile interacts with other quarks.
The direct contribution is evidently given by the formula analogous to
(\ref{12}). We shall limit ourselves with the inclusive cross-section
integrated over the transverse momentum $I_A(x)=xd\sigma/dx$. The direct
part is then
\beq
I_A^{dir}(x)=\int_{x}(d\xi /\xi)D(\xi )h(x/\xi )
\label{30}
\eeq
where again $D(\xi)/\xi$ is the probability
defined in the preceding subsection
to find in the nucleus a cumulative quark with
scaling variable $\xi$, $h(x/\xi)$ is the
inclusive cross-section with this quark as a target,
$x =k_{+}/p_{+}\simeq k_z/p_z$
and $\ksi =k'_{+}/p_{+}\simeq k'_z/p_z$.
Again in the cumulative region $1<x<\ksi<A$.

%Fig.2

The spectator part has a much more complicated structure.
We have analysed its contribution in details in \cite{NPB,dub95}.
We have found that
near the thresholds significant cancellations occur in this part,
which lead to the domination of the diagrams, with
{\it each} donor quark
interacting with the projectile .
As a result the contribution of the spectator mechanism
{\it has not} such a simple convolutional form (\ref{30})
as in the direct case.
With the quasi-eikonal parametrization of the partonic amplitude chosen to
account for diffraction, we found \cite{NPB,dub95} for the contribution
of the spectator mechanism
to the cumulative particle production
in the case when $n-1$ donors transfer
their momenta to the active quark:
\beq
I_A^{n,sp}(x)=D_n(x)C_P^{-1}
\int d^{2}B [4\pi m^{2}j(B)]^{n-1}
\label{Isp}
\eeq
with
\beq
j(B)=C_P \nu_P \int d^{2}b d^{2}b' \lambda^{2}(b) \rho_P (b')
|a(B+b'-b)-a(B+b')|^{2}
\label{jsp}
\eeq
where
\beq
\lambda (b)=K_{0}(mb)/2\pi
\eeq
Here $\nu_P$ is the mean number of partons in the projectile,
$\rho_P$ is the one parton distribution normalized to unity
and $C_P$ is the quasi-eikonal factor for the projectile.
For example, for NN-interaction
$C_N^2=1+\sigma_{NN}^{dif}/\sigma_{NN}^{el}$.
The $a$ is the partonic amplitude of the interaction between a parton of the
projectile and a donor quark from the nucleus.

\section{The cumulative quark production probability}

We see that both the nuclear structure function (\ref{12}) and
the particle production rate (\ref{30},\ref{Isp}) in cumulative region
are determined by
$D(\ksi)/\ksi$ - function which
has the meaning of the probability to find in the
nucleus a quark with the scaling variable $\ksi > 1$.

%Fig.3
\begin{figure}

\unitlength=0.5mm
\linethickness{0.4pt}
\begin{picture}(127.00,101.00)(-50,20)
\put(45.00,95.00){\line(1,0){30.00}}
\put(45.00,86.00){\line(1,0){30.00}}
\put(45.00,77.00){\line(1,0){30.00}}
\put(45.00,59.00){\line(1,0){30.00}}
\put(45.00,68.00){\line(1,0){30.00}}
\put(45.00,41.00){\line(1,0){55.00}}
\put(45.00,50.00){\line(1,0){55.00}}
\put(45.00,23.00){\line(1,0){55.00}}
\put(45.00,32.00){\line(1,0){55.00}}
\put(37.50,59.00){\oval(15.00,84.00)[]}
\put(75.00,56.00){\framebox(36.00,42.00)[cc]{$B_n$}}
\put(30.00,60.00){\line(-1,0){15.00}}
\put(15.00,60.00){\line(0,0){0.00}}
\put(30.00,58.00){\line(-1,0){15.00}}
\put(12.00,59.00){\makebox(0,0)[rc]{$A$}}
\put(111.00,86.00){\line(1,0){13.00}}
\put(111.00,77.00){\line(1,0){13.00}}
\put(111.00,68.00){\line(1,0){13.00}}
\put(111.00,59.00){\line(1,0){13.00}}
\put(111.00,95.00){\line(1,0){13.00}}
\put(127.00,95.00){\makebox(0,0)[lc]{$x_1=\xi\rightarrow nx_0$}}
\put(127.00,86.00){\makebox(0,0)[lc]{$x_2\rightarrow 0$}}
\put(127.00,59.00){\makebox(0,0)[lc]{$x_n\rightarrow 0$}}
\put(60.00,61.00){\makebox(0,0)[cb]{$x_0$}}
\put(60.00,88.00){\makebox(0,0)[cb]{$x_0$}}
\put(60.00,97.00){\makebox(0,0)[cb]{$x_0$}}
\end{picture}

\caption{Formation of the hard parton component
of the flucton in the nucleus.}
\label{f3}
\end{figure}
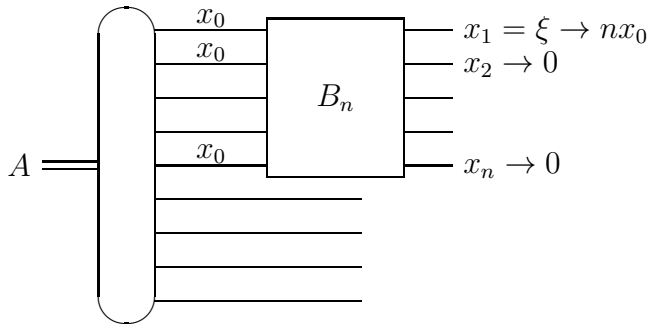

Study the general case when $n-1$ quarks aquire
their momenta equal to zero as a result of interquark interactions,
giving their momentum to the active quark (Fig.~\ref{f3}).
The scaling variables for quarks are defined as
 $x_i=k_{i+}/p_{+}$.
 According to our assumption initially in the nucleus
all $x_{i}\simeq x_{0}$ and the distribution in $x_{i}$ is sharp. The
momentum
of the active quark is denoted as $k=k_1$ and its scaling variable
 is $\ksi =x_1$. Evidently with $n-1$ momentum exchanges its
maximal value is $nx_{0}$ and when $x_1=\xi\rightarrow nx_0$ then
all $x_i\rightarrow 0$ for $i=2,3,...,n$.
Particulaly on this fact the use of the perturbation theory is based
\cite{14}.

The block $B_n$ describes the formation of the hard parton component
of the flucton as a result of $n-1$ interquark interactions.
Due to the quark counting rules \cite{11,12} one has
$D_n(\xi)\sim (nx_0-\xi)^{2n-3}$ when $\xi\rightarrow nx_0$.
But as we have mentioned in the introduction
for to calculate the cumulative quark production probability
one needs besides to know the coefficients
of the quark counting rules behaviour, which determine the relative weights
with which  contributions with different $n$ combine
to give the total probability, as
well as the overall  normalization.
So unlike \cite{8} we are trying to calculate
theoretically the structure function of the flucton
at least its hard component.

\begin{figure}

\unitlength=0.5mm
\linethickness{0.4pt}
\begin{picture}(73.00,60.00)(-80,10)
\put(10.00,10.00){\line(1,0){60.00}}
\put(10.00,20.00){\line(1,0){60.00}}
\put(10.00,50.00){\line(1,0){60.00}}
\put(10.00,60.00){\line(1,0){60.00}}
\put(10.00,30.00){\line(1,0){60.00}}
\put(10.00,40.00){\line(1,0){60.00}}
\put(7.00,10.00){\makebox(0,0)[rc]{$x_0$}}
\put(7.00,50.00){\makebox(0,0)[rc]{$x_0$}}
\put(7.00,60.00){\makebox(0,0)[rc]{$x_0$}}
\put(7.00,40.00){\makebox(0,0)[rc]{$x_0$}}
\put(73.00,60.00){\makebox(0,0)[lc]{$\lx1$}}
\put(73.00,50.00){\makebox(0,0)[lc]{$\x 2$}}
\put(73.00,40.00){\makebox(0,0)[lc]{$\x 3$}}
\put(73.00,10.00){\makebox(0,0)[lc]{$\x n$}}
\v{20}{50}{60}{10}
\v{30}{40}{60}{20}
\v{40}{30}{60}{30}
\v{50}{20}{60}{40}
\v{60}{10}{60}{50}
\end{picture}

\caption{An example of the diagram which contribute to $B_n$.}
\label{f4}
\end{figure}
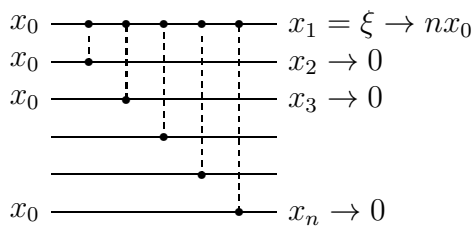

In the perturbative approach we find Feynman diagrams
for the $B_n$ of the type shown in Fig.~\ref{f4}.
%Fig.4
The shown diagram is only an example. It contains a particular set of
interquark interactions, in which donors successively give their
momentum to the active quark and become soft.
However other type of interquark interactions are also possible: say,
the $n-1$ donors may give their momenta successively to each other, the
last giving all the accumulated momentum to the active quark
(see second diagram in \F 6).
On these diagrams dash vertical lines refer to one gluon
exchange. One can choose the Coulomb gauge in which the transversal part is
damped at low $x_{i}$ \cite{14} and the dominating Coulomb part  depends
only on the scaling variables of the participant quarks before ($x_{1},
x_{2}$) and after ($x'_{1},x'_{2}$) the interaction
(see \F 7). If we neglect the
colour then
\beq
V=4\pi\alpha (x_{1}+x'_{1})(x_{2}+x'_{2})/ (x_{1}-x'_{1})^{2}
=4\pi\alpha \tilde{V}
\label{3}
\eeq
with $\alpha$ the interaction constant. To approximately take the colour
into account we average over the quark colour variables, which
introduces a factor $\sqrt{2}/3$ into (\ref{3}).

We are using the time ordered pertubation theory
in the light cone variables.
In the energy denominators of the intermediate states the
contributions from soft final quarks dominate
as their "energeis" $E_i=(m^2+\k^2_{i\perp})/x_i$ rise with $\x i$
for $i=2,3,...,n$.
Thus the contribution of the diagram in \F 4 will be proportional
to $[E_2(E_2+E_3)...(E_2+...+E_n)]^{-1}$ or after symmetrization
on $x_2,...,x_n$ to $[E_2 E_3...E_n]^{-1}$.

The advantage of the time ordered pertubation theory
is that the energy denominators have the same form for any diagram
in the threshold limit and coincied after symmetrization on $x_2,...,x_n$.
To illustrate, we have presented in \F 5 one Feynman diagram as the sum of
two time ordered diagrams.
%Fig.5
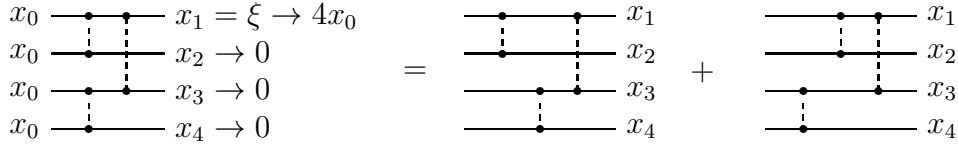
\begin{figure}

\unitlength=0.5mm
\linethickness{0.4pt}
\begin{picture}(203.00,40.00)(-30,10)
\put(10.00,10.00){\line(1,0){30.00}}
\put(10.00,20.00){\line(1,0){30.00}}
\put(10.00,30.00){\line(1,0){30.00}}
\put(10.00,40.00){\line(1,0){30.00}}
\put(7.00,10.00){\makebox(0,0)[rc]{$x_0$}}
\put(7.00,20.00){\makebox(0,0)[rc]{$x_0$}}
\put(7.00,30.00){\makebox(0,0)[rc]{$x_0$}}
\put(7.00,40.00){\makebox(0,0)[rc]{$x_0$}}
\put(43.00,40.00){\makebox(0,0)[lc]{$x_1=\xi\rightarrow 4x_0$}}
\put(43.00,30.00){\makebox(0,0)[lc]{$\x2$}}
\put(43.00,20.00){\makebox(0,0)[lc]{$\x3$}}
\put(43.00,10.00){\makebox(0,0)[lc]{$\x4$}}

\put(120.00,10.00){\line(1,0){40.00}}
\put(120.00,20.00){\line(1,0){40.00}}
\put(120.00,30.00){\line(1,0){40.00}}
\put(120.00,40.00){\line(1,0){40.00}}
\put(163.00,40.00){\makebox(0,0)[lc]{$x_1$}}
\put(163.00,30.00){\makebox(0,0)[lc]{$x_2$}}
\put(163.00,20.00){\makebox(0,0)[lc]{$x_3$}}
\put(163.00,10.00){\makebox(0,0)[lc]{$x_4$}}

\put(110.00,25.00){\makebox(0,0)[rc]{=}}
\put(180.00,25.00){\makebox(0,0)[lc]{+}}

\put(200.00,10.00){\line(1,0){40.00}}
\put(200.00,20.00){\line(1,0){40.00}}
\put(200.00,30.00){\line(1,0){40.00}}
\put(200.00,40.00){\line(1,0){40.00}}
\put(243.00,40.00){\makebox(0,0)[lc]{$x_1$}}
\put(243.00,30.00){\makebox(0,0)[lc]{$x_2$}}
\put(243.00,20.00){\makebox(0,0)[lc]{$x_3$}}
\put(243.00,10.00){\makebox(0,0)[lc]{$x_4$}}

\v{20}{10}{20}{10}
\v{20}{30}{40}{10}
\v{30}{20}{40}{20}
\v{130}{30}{40}{10}
\v{140}{10}{20}{10}
\v{150}{20}{40}{20}
\v{210}{10}{20}{10}
\v{220}{30}{40}{10}
\v{230}{20}{40}{20}

\end{picture}

\caption{Decomposition of the Feynman diagram on the time ordered diagrams.}
\label{f5}
\end{figure}
The Feynman diagram is proportional
to $[E_2 E_4 (E_2+E_3+E_4)]^{-1}$ and the time ordered diagrams
are proportional to $[E_2(E_2+E_4)(E_2+E_4+E_3)]^{-1}$  and
to $[E_4(E_4+E_2)(E_4+E_2+E_3)]^{-1}$ respectively
in the $\xi\rightarrow 4x_0$ limit.
After symmetrization on $x_2,...,x_n$ the contributions of the last
two diagrams both become proportional to $[E_2 E_3 E_4]^{-1}$.
Moreover in this case they are equal which leads to the possibility
to take into account only one of them multiplied by the "time ordering"
factor 2.

As result the $B_n$ can be presented in the form
\beq
B_n=X_n\frac{(4\pi\alpha)^{n-1}}{(x_0)^{n-2}}\frac 1 {E_2 E_3...E_n}
\label{B_n}
\eeq
The denominator leads after integration over $x_2,...,x_n$
to the quark counting rule behavior for the cumulative quark
production rate from the $n$-quark flucton.
But our aim is to calculate the coefficient in this behavior,
which determines the relative weights
with which contributions on different thresholds
combine to give the total probability.
For to do this we have to calculate the sum of all possible Feynman
diagrams of the type shown in \F 6. In (\ref{B_n}) the $X_n$ denotes
this sum after the extraction of the common factor
$(4\pi\alpha)^{n-1} (x_0)^{2-n} [E_2 E_3...E_n]^{-1}$.

%Fig.6
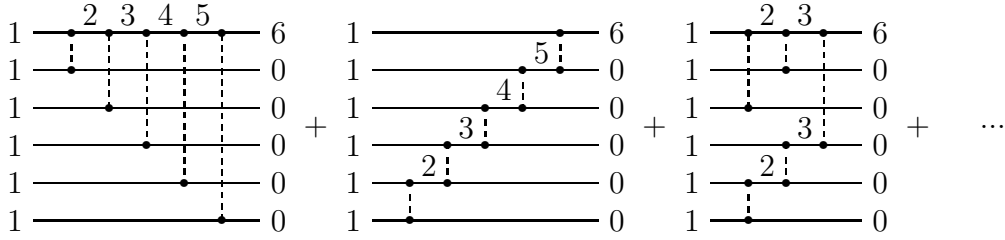
\begin{figure}

\unitlength=0.5mm
\linethickness{0.4pt}
\begin{picture}(200,62)(-20,10)
\put(0,0){
\begin{picture}(85,62)
\multiput(10,10)(0,10){6}{\line(1,0){60}}
\multiput(7,10)(0,10){6}{\makebox(0,0)[rc]{$1$}}
\multiput(73,10)(0,10){5}{\makebox(0,0)[lc]{$0$}}
\put(73,60){\makebox(0,0)[lc]{$6$}}
\v{20}{50}{60}{10}
\v{30}{40}{60}{20}
\v{40}{30}{60}{30}
\v{50}{20}{60}{40}
\v{60}{10}{60}{50}
\put(25,62){\makebox(0,0)[cb]{$2$}}
\put(35,62){\makebox(0,0)[cb]{$3$}}
\put(45,62){\makebox(0,0)[cb]{$4$}}
\put(55,62){\makebox(0,0)[cb]{$5$}}
\put(85,35){\makebox(0,0)[cc]{+}}
\end{picture}
}
\put(90,0){
\begin{picture}(85,62)
\multiput(10,10)(0,10){6}{\line(1,0){60}}
\multiput(7,10)(0,10){6}{\makebox(0,0)[rc]{$1$}}
\multiput(73,10)(0,10){5}{\makebox(0,0)[lc]{$0$}}
\put(73,60){\makebox(0,0)[lc]{$6$}}
\v{20}{10}{20}{10}
\v{30}{20}{30}{10}
\v{40}{30}{40}{10}
\v{50}{40}{50}{10}
\v{60}{50}{60}{10}
\put(25,22){\makebox(0,0)[cb]{$2$}}
\put(35,32){\makebox(0,0)[cb]{$3$}}
\put(45,42){\makebox(0,0)[cb]{$4$}}
\put(55,52){\makebox(0,0)[cb]{$5$}}
\put(85,35){\makebox(0,0)[cc]{+}}
\end{picture}
}
\put(180,0){
\begin{picture}(85,62)
\multiput(10,10)(0,10){6}{\line(1,0){40}}
\multiput(7,10)(0,10){6}{\makebox(0,0)[rc]{$1$}}
\multiput(53,10)(0,10){5}{\makebox(0,0)[lc]{$0$}}
\put(53,60){\makebox(0,0)[lc]{$6$}}
\v{20}{10}{20}{10}
\v{30}{20}{30}{10}
\v{40}{30}{60}{30}
\v{20}{40}{60}{20}
\v{30}{50}{60}{10}
\put(25,22){\makebox(0,0)[cb]{$2$}}
\put(35,32){\makebox(0,0)[cb]{$3$}}
\put(25,62){\makebox(0,0)[cb]{$2$}}
\put(35,62){\makebox(0,0)[cb]{$3$}}
\put(65,35){\makebox(0,0)[cc]{+}}
\put(85,35){\makebox(0,0)[cc]{...}}
\end{picture}
}

\end{picture}
\caption{The diagrams which contribute to $X_n$ for $n=6$ as an example.
All $x_i$ are in units $x_0$.}
\label{f6}
\end{figure}

%Fig.7
\begin{figure}
\unitlength=0.5mm
\linethickness{0.4pt}
\begin{picture}(200,20)(-30,10)

\put(0,0){
\begin{picture}(60,20)
\multiput(10,10)(0,10){2}{\line(1,0){20}}
\put(7,10){\makebox(0,0)[rc]{$x_2$}}
\put(7,20){\makebox(0,0)[rc]{$x_1$}}
\put(33,10){\makebox(0,0)[lc]{$x'_2$}}
\put(33,20){\makebox(0,0)[lc]{$x'_1$}}
\put(50,15){\makebox(0,0)[cc]{=}}
\put(60,15){\makebox(0,0)[lc]{${\displaystyle
\frac{(x_1+x'_1)(x_2+x'_2)}{(x_1-x'_1)^2}\ ;}$}}
\v{20}{10}{20}{10}
\end{picture}
}

\put(130,0){
\begin{picture}(60,20)
\put(15,15){\line(1,0){20}}
\put(20,15){\circle*{2.00}}
\put(30,15){\circle*{2.00}}
\put(25,17){\makebox(0,0)[cb]{$x_i$}}
\put(45,15){\makebox(0,0)[cc]{=}}
\put(55,15){\makebox(0,0)[lc]{${\displaystyle \frac1{x_i}}$}}
\end{picture}
}

\end{picture}

\caption{The diagram rules for $X_n$.}
\label{f7}
\end{figure}

\begin{figure}

\unitlength=0.5mm
\linethickness{0.4pt}
\begin{picture}(200,62)(-20,10)
\put(0,0){
\begin{picture}(95,63)

\put(20,7){\framebox(50,56)[cc]{$X_n$}}

\multiput(10,10)(0,10){6}{\line(1,0){10}}
\multiput(70,10)(0,10){6}{\line(1,0){10}}

\multiput(7,10)(0,10){6}{\makebox(0,0)[rc]{$1$}}
\multiput(83,10)(0,10){5}{\makebox(0,0)[lc]{$0$}}
\put(83,60){\makebox(0,0)[lc]{$n$}}

\put(115,35){\makebox(0,0)[cc]{=}}
\end{picture}
}

\put(160,0){
\begin{picture}(85,63)

\put(20,8){\framebox(30,24)[cc]{$X_{n-k}$}}
\put(20,38){\framebox(30,24)[cc]{$X_{k}$}}

\multiput(10,10)(0,10){6}{\line(1,0){10}}
\multiput(50,10)(0,10){6}{\line(1,0){30}}

\multiput(7,10)(0,10){6}{\makebox(0,0)[rc]{$1$}}
\multiput(83,10)(0,10){5}{\makebox(0,0)[lc]{$0$}}
\put(83,60){\makebox(0,0)[lc]{$n$}}

\put(0,35){\makebox(0,0)[rc]{${\displaystyle \sum_{k=1}^{n-1} C^{k-1}_{n-2}}$}}
\put(60,32){\makebox(0,0)[cb]{$n\!-\!k$}}
\put(60,62){\makebox(0,0)[cb]{$k$}}

\v{70}{30}{60}{30}

\end{picture}
}

\end{picture}

\caption{The diagrammatic recurrency relation for $X_n$.}
\label{f8}
\end{figure}
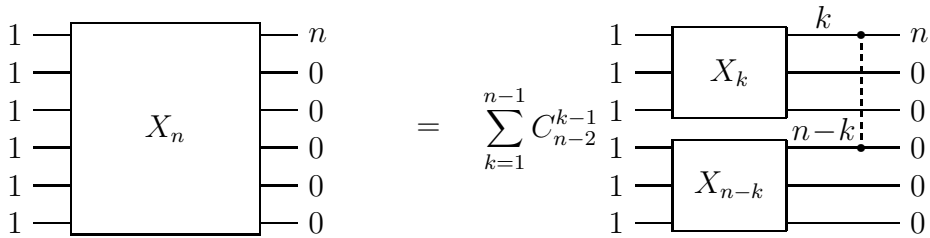

After this extraction one can go to the limit
$x_1=\xi=nx_0$ and $x_i=0$ for $i=2,...,n$ in the rest of the diagram,
which leads to the diagram rules in \F 7 for $X_n$.
All $x_i$ in Fig.~\ref{f6}-\ref{f8} are in units $x_0$.
We have also to multiply each diagram on the proper "time ordering"
factor for to take into account the different time ordering
possibilities for gluon exchanges.

The main problem here is that the number of these diagrams
grows rapidly with $n$, as $n!$. For to sum all these diagrams
we derive a recurrency relation. The derivation is based on the
important observation that
only diagrams with all intermediate $x$
greater than zero should be retained. Separating the last exchange, which
raises the momentum of the active quark up to its final value we present
$X_n$ as a sum over $k=1,2,...,n-1$ of the diagrams shown in \F 8, where it
has been assumed that $X_1=1$ by definition.
%Fig.8
Times of $k-1$ gluon
exchanges in the upper blob $X_k$ relative to those in the lower blob
$X_{n-k}$ may be arbitrary, which leads to a time ordering factor
$C_{n-2}^{k-1}$.
Taking into account the diagram rules in \F 7 and
denoting $f_n\equiv X_n/n!$ we find that \F 8
translates into a recurrency relation for $f_n$:
\beq
f_{n}=\frac 1 {n(n-1)}\sum_{k=1}^{n-1}\frac{n+k}{n-k}f_{k}f_{n-k}
\label{24}
\eeq
with the initial condition $f_{1}=1$.
The recurrency relation (\ref{24}) enables easy calculate $f_n$
for an arbitrary $n$ starting from $f_{1}=1$.
For large $n$ (\ref{24}) evidently admits
asymptotical solutions of the form
\beq
f_{n}\simeq [(6/5)n+o(n)]\exp (-an)
\label{25}
\eeq
where $a$ is arbitrary. Numerical studies reveal that with $f_{1}=1$
\[ a=0.24421...\]
and also show that the asymptotical expression (\ref{25})
appproximates the true
solution quite well starting from $n=3$, i.e. for all physically
interesting values.

%%%%%%%%%%%%%%%%%%%%%%%%%%%%%%%%%%%%%%%%%%%%%%%%%%%%%%%%

So we have calculated the $X_n=f_n n!$ and consequently the $B_n$
(\ref{B_n}) which describes the formation of the hard parton component
of the flucton. Now from diagram in \F 3 we can calculate
the $D(\xi)/\xi$ - the probability
to find in the nucleus a hard cumulative quark with
scaling variable $\xi$.

The block $B_n$ is not dependent on the momentum variables
of the initial quarks. As the consequence we find that
all donor quarks
have their coordinates relative to the active one equal to zero.

Now assume that on
the average quarks are homogeneously distributed inside nucleons (three
in each of them).
In this picture the
probability $w_{n}^{(q)}$ to find $n-1$ quarks within the distance $R_{1}$
from the active one, where $R_{1}$ is the nucleon radius, is equal to
the probability to find within the same distance $\n$ nucleons,
where $\n = [(n-1)/3]+1 $
(i.e the entire part of $(n-1)/3$ plus one nucleons):
\beq
w_{n}^{(q)}=w_{\n}^{(N)}
\label{17}
\eeq
If the nucleons, in their turn, are homogeneously distributed inside the
nucleus of the finite volume $V_{A}=AV_{0}$ then the latter probability
will be given by
\beq
w_{\n}^{(N)}=C^{\n-1}_{A-1}(V_{1}/V_{A})^{\n-1}
\label{18}
\eeq
where $V_{1}$ is the nucleon's volume.
Actually, as well-known, this approximation is rather poor due to the
rise of the nuclear density towerds its center. To account for this fact
we introduce a factor $q_{n}$ to be calculated from a realistic
nuclear distribution. In particular for the Saxon-Woods distribution
\beq
\rho(r)=\rho_{0}/(1+\exp (r-R_{A})/r_{1}))
\label{19}
\eeq
with $R_{A}=A^{1/3}\,1.17\ fm$ the nuclear radius and $r_{1}=0.51\ fm$,
for $A>>1$
\beq
q_{\n}=1-(3r_{1}/R_{A})\sum_{k=1}^{\n-1}(1/k)=1-1.31\,A^{-1/3}
\sum_{k=1}^{\n-1}(1/k)
\label{20}
\eeq

An explicit calculation gives for the
cumulative quark distribution
\beq
D_{n}(\ksi )=3AC^{\n-1}_{A-1}X_{n}^{2}
\left(\frac{\alpha^{2}}{9m^{3}V_{1}}\right)^{n-1}q_{\n}
\left(\frac{V_{1}}{V_{A}}\right)^{\n-1}
\frac{\theta(\Delta_{n})\Delta_{n}^{2n-3}}{(2n-3)!}
\label{23}
\eeq
where $\n-1=[(n-1)/3]$ is the "cumulative number"
(i.e. the number of nucleons in the flucton $\n$ minus one) and
$\Delta_{n}=(nx_0-\xi)/x_0$.
All nontrivial $A$ dependence of the distribution $D$ is contained in
the nuclear factor $q_{\n}$. It   rises with $A$ approaching its
asymptotic value 1. The rise is quicker for higher $\n$.
Studies of  the cumulative
particle production rate in the nucleon picture, where the same $A$-
dependence emerges \cite{6}, revealed that it
agrees rather well with the experimental
observations.

With (\ref{25}) we finally find
\beq
D_{n}(\xi)=\frac{216}{25}\frac{A}{(\n-1)!}q_{\n}
\left(\frac{V_{1}}{V_{0}}\right)^{\n-1}
(n-1)n^{9/2}\gamma^{n-1}
\theta(\Delta_{n})\Delta_{n}^{2n-3}
\label{26}
\eeq
where,
\beq
\gamma=\alpha^{2}/36m^{3}V_{1}\exp 2a
\label{27}
\eeq
and we have assumed $A>>1$. The behavior of $D_{n}(\xi)$ near the
threshold $\Delta_{n}=0$ is determined by the factor $\Delta_{n}^{2n-3}$,
which corresponds to the well-known quark counting rules \cite{11}. The
magnitude of the coefficient is determined mainly by the factor
$\gamma^{n-1}$. As we shall see later, the values of $\gamma$
extracted from the known experimental data on nuclear
structure functions in the cumulative region
 result quite small.
Because of this the power factor $\gamma^{n-1}$
 ensures that the coefficient in the $\xi$ dependence of (\ref{26})
 drops very fast with the growth of $n$. Down from the treshold the
contribution $D_{n}(\xi)$ rises as a power.  This rise has evidently to
stop at some distance from the threshold where our approximations
become invalid and terms with higher powers of $\Delta_{n}$ become
important. We expect, however, that the overall order of the contribution
remains to be governed by the small factor $\gamma^{n-1}$ coming
 from $n-1$ gluon exchanges.

As a result, due to the smallness of $\gamma$, the contribution from
diagrams
 with a given $n$ will always
dominate over those with greater $n$'s, as soon as we slightly move
down from the threshold point $\ksi=nx_{0}$.
 Therfore we may safely
assume that $D_{n}(\ksi)$ given by (\ref{23}) or (\ref{26}) represents
 the true quark distribution
for $\ksi$ starting  from the threshold $\ksi=nx_{0}$ down to
the point
$\ksi=(n-1)x_{0}$, which is the threshold for the diagrams with $n-2$
donors, i. e. in the region
\beq
(n-1)x_{0}\leq\ksi <nx_{0},\ \ x_{0}=1/3
\label{28}
\eeq
Alternativelely, to avoid discontinuities, we may take the total
distribution $D(\xi)$  as a sum of  $D_{n}(\xi)$ over all possible
$n$'s with an additional assumption that $\Delta_{n}=1$ for
$\xi$ smaller than $(n-1)x_{0}$, which means freezing the contribution
$D_{n}$ at the threshold for the contribution $D_{n-1}$.
As a result we have the roughly exponential
behaviour of $D(\xi)$ in the cumulative region,
governed by an effective coupling constant,
which depends on the QCD coupling constant and quark mass.

\section{Numerical calculations, discussion
and the comparison to experimental data}

Both convolutional formulas (\ref{12},\ref{30})
lead to the same $x$
dependence for the structure function
$F_{2}^{A}(x)$ and for the direct mechanism production rate $I_A^{dir}(x)$
in the region $x>1$. It is roughly an exponential in $x$
\beq
I_A^{dir}(x) \sim F_{2}^{A}(x) \sim \exp(-b_{0}x)
\label{dir}
\eeq
where the slope $b_{0}$ is determined by the QCD coupling constant and
quark mass.

The spectator mechanism (\ref{Isp}) besides involves interactions
between partons of the projectile and target nucleus. It was shown in
\cite{NPB} that each donor quark has to interact with the projectile.
As a result the spectator contribution $I_A^{sp}(x)$
also behaves in $x$ as an exponential but with a different slope
\beq
I_A^{sp}(x) \sim \exp (-b_{s}x)
\label{sp}
\eeq
the slope $b_{s}$ depending not only on the QCD coupling constant and
quark mass but also on the partonic amplitude.

%++++++++++++++++++++++

In \cite{dub95} we have found that final state interactions
cancel the leading terms in the direct
contribution, so that it becomes much less than the spectator one. Then
particle production in the cumulative region indeed goes predominantly
via the spectator mechanism.
To be able to explain the experimental slope
we have studied in \cite{dub95}
the quasi-eikonal parametrization for the partonic amplitude.
The spectator slope $b_{s}$ results very sensitive to the
magnitude of the hadronic diffractive cross-sections.
We have chosen
the maximum possible value for the diffractive parameter
for the case of the nucleon as a projectile $C_P=C_N=1.5$ (see (\ref{jsp})).
Then the quasi-eikonal parametrization leads to values of
the partonic amplitude $a$ (\ref{jsp}) considerably
larger than without diffraction because of a stronger screening effect
introduced by diffractive states.

The results of the calculations are shown in Fig.~9 for the
cumulative charged pion production on $^{181}$Ta,
respectively. For comparison the corresponding nuclear structure
function $F_{2}^A(x)$ at $x>1$, is also shown, as
well as the available experimental data from \cite{Baldin82}-\cite{Benven94}.

One clearly observes that the spectator mechanism,
with a quasi-eikonal para\-met\-ri\-za\-tion
of the partonic amplitude chosen to account for diffraction, leads to a
considerably smaller slope of the production spectra ($b_{s}\sim 7\div 9$)
(\ref{sp}) compared to the slope of the structure function (\ref{dir}) in the
region $x>1$ ($b_{0}\sim 16$), in a good agreement with experimental
data.

Thus our model correctly predicts two differents exponential in the
cumulative particle production and
in the nuclear structure function at $x>1$.
The difference is due to additional multiple interactions between
projectile and target which enter the spectator mechanism for the
cumulative production.

A more phenomenological attempt to explain smaller slopes for
the cumulative particle production is made in \cite{Efremov94},
where the existence
of multiquark clusters in the nucleus  and their properties are
postulated and the quark-gluon string model formalism of \cite{Kaidalov82}
is used to calculate production rates.

Note that at a nucleon level of analysis a similar approach
to a description of cumulative phenomena was suggested
in our paper \cite{YF80} in which
to find the cumulative particle production rates
%cross-section
we used the fragmentation functions of multi-nucleon fluctons
calculated in the framework of
%multi-reggeon
Regge approach.
In a sense, the calculations of the probabilities "to slow"
the quark system with some definite quantum numbers
usually discussed in the quark-gluon string model approach
\cite{Kaidalov82,Kaidalov80,Kaid_Pisk85}
%is equivalent
is close
to the idea of counting of the quantum numbers
transferring from the flucton fragmentation region
to the central one
in triple reggeon like approach
%in the approach which is like to the triple reggeon one
\cite{YF80}.
In the last case these quantum numbers
transferring in $t$-chanel determine
the number and the type of Regge trajectories
which have to be included
in the "shoulders" of the triple like reggeon diagram.

To calculate the hadron production rates
in the cumulative region we need also to analyze
the process of fragmentation of one (or several) fast quark(s) into
hadrons of different flavors.
In principle
the different versions of hadronization mechanisms of the produced
fast cumulative quarks into colorless states (hadrons)
can be suggested.
Apparently that in the case of the production of cumulative protons
the hadronization through the coalescence of three cumulative quarks
is favorable than the usual hadronization through one cumulative quark
fragmentation into proton.
This leads to the higher production rates for the protons
compared with the pions
but approximately with the same slopes of the protons and mesons spectra
in agreement with experiment.

The idea that in QCD a quark can hadronize by coalescing with
a comoving spectator parton was suggested in the paper
\cite{BGS87}.
It was used later for the
description of the fragmentation of protons and pions
into charm and beauty hadrons at large $x$
\cite{Vogt92,Vogt95}.
It was shown that the coalescence or recombination of
one or both intrinsic charm quarks with spectator valence
quarks of the Fock state leads in a natural way to
leading charm and beauty production.

The conception of quark fusion
for the description of hadronic fragmentation into
$\pi ^{\pm}$ and $K^{\pm}$ mesons was also used in the papers
\cite{Meng94,Meng_96}.
It was shown that the valence quarks of the projectile hadron
play a dominating role in meson production
in the projectile fragmentation region
and
while the $x$-distribution for $\pi ^+$, $\pi^-$ and
$K^+$ are the convolutions of the $u_v$, $d_v$ and
$u_v$ valence quarks with those of the corresponding
antiseaquarks $\bar d_s$, $\bar u_s$ and $\bar s_s$,
the $x$-distribution for $K^-$ is that of the
a sea quark $s_s$ and an antiseaquark $\bar d_s$.
The spin properties of these fragmentation processes
have been also studied within framework of this approach
\cite{Meng93}-\cite{Meng96}.
In particular it has been shown that the observed
left-right asymmetry can be readily described in the
framework of a relativistic quark model in which
the observed $\pi ^+$ and $\pi ^-$ are respectively
the fusion products of the valence quarks
$u_v$ and $d_v$ of $p(\uparrow )$ and
antiseaquarks $\bar d_s$ and $\bar u_s$.

Note that both
the intrinsic mechanism of the cumulative quark production
when the quarks of several nucleons concentrated
in one nuclear flucton transfer their
longitudinal momenta to the distinguished quark
and
the hadronization through the cumulative quarks coalescence
break the QCD factorization theorem.

This work is supported by the Russian Foundation for Fundamental
Research, Grant No. 94-02-06024-a.

%\vskip 2 cm
\newpage

%Fig.9
\begin{figure}[tp]
\centerline{\psfig{bbllx=4cm,bblly=5cm,bburx=17cm,bbury=
18cm,file=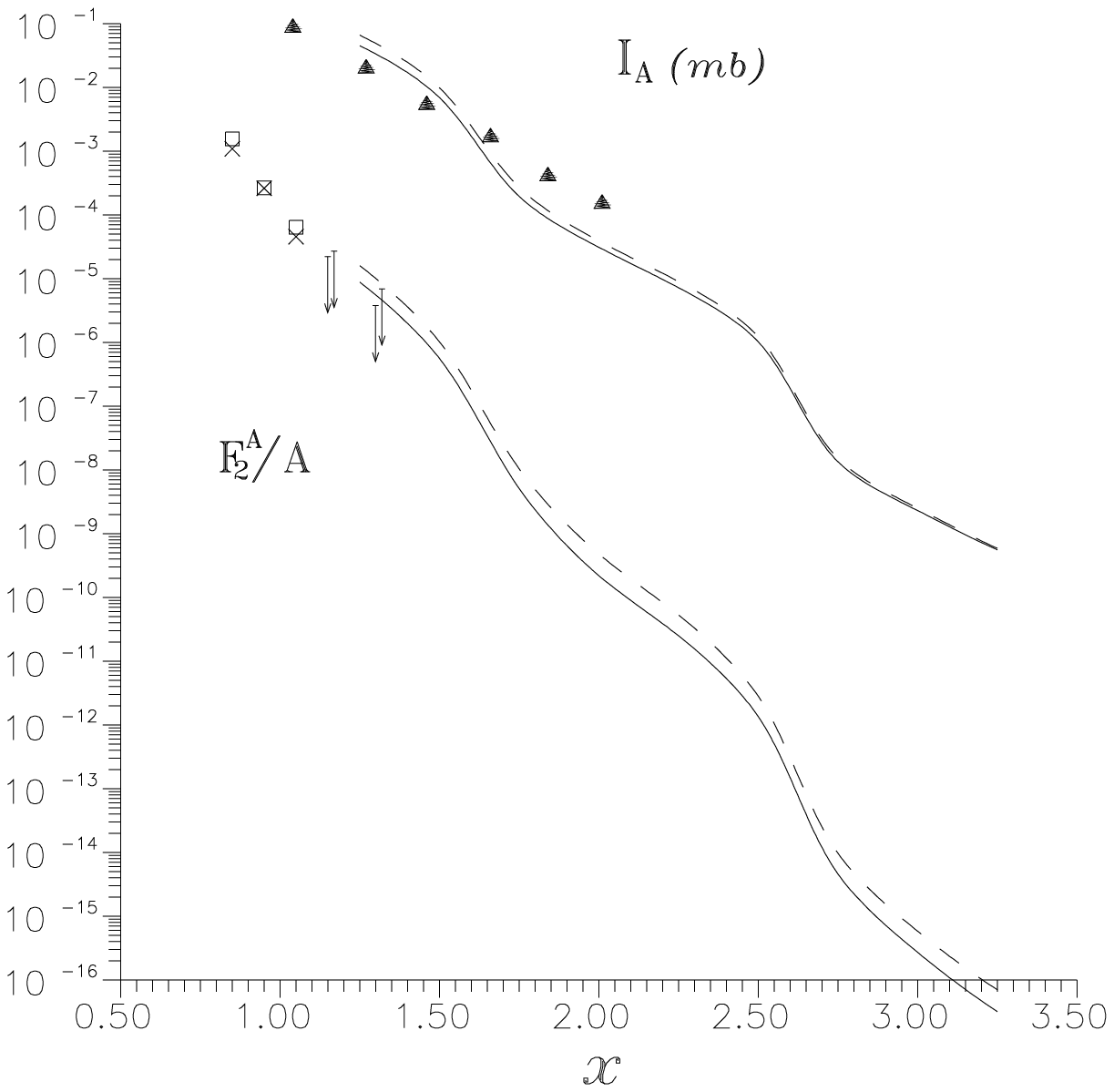,height=13cm,width=13cm,rheight=13cm}}
\caption[9]{\label{fig_9}
$I_A=\frac{xd\sigma}{Adx}(mb)$
is the calculated inclusive cross-section (per nucleon)
for cumulative charged pion
production on $^{181}{\rm Ta}$
at  $\sqrt{s}=23.5\ GeV$ (solid curve) and
$1800\ GeV$ (dashed one).
$\Delta$ - the experimental data \cite{23} on the
cumulative charged pion production on
$^{181}{\rm Ta}$
by $400\ GeV$ incident proton beam.
$F^A_2/A$ is
the calculated nuclear structure function for the
$^{181}{\rm Ta}$ at $Q^2=50\ GeV^2$ (dashed curve) and
$^{12}{\rm C}$ at $Q^2=100\ GeV^2$ (solid curve).
$\Box$ and $\times$ - the experimental data \cite{Benven94} on the
$^{12}{\rm C}$ structure function
at $Q^2=61\ GeV^2$ and  $150\ GeV^2$, respectively.
}
\end{figure}

\end{document}